\documentclass[journal]{IEEEtran}
\usepackage{cite}
\usepackage{graphicx}
\usepackage{amsmath}
\usepackage{algorithmic}
\usepackage{array}
\usepackage{url}

\begin{document}

\title{Baba is Y'all 2.0: \\Design and Investigation of a \\Collaborative Mixed-Initiative System}

\author{M Charity,
        Isha Dave,
        Ahmed Khalifa,
        and~Julian Togelius%
\thanks{M Charity, Isha Dave, and Julian Togelius are with New York University in the Game Innovation Lab.}
\thanks{Ahmed Khalifa is with University of Malta and Modl.ai.}
}


\maketitle

\begin{abstract}
This paper describes a new version of the mixed-initiative collaborative level designing system: Baba is Y'all, as well as the results of a user study on the system. Baba is Y'all is a prototype for AI-assisted game design in collaboration with others. The updated version includes a more user-friendly interface, a better level-evolver and recommendation system, and extended site features. The system was evaluated via a user study where participants were required to play a previously submitted level from the site and then create their own levels using the editor. They reported on their individual process creating the level and their overall experience interacting with the site. The results have shown both the benefits and limitations of this mixed-initiative system and how it can help with creating a diversity of `Baba is You' levels that are both human and AI designed while maintaining their quality.
\end{abstract}

\begin{IEEEkeywords}
Mixed-Initiative, Level Design, PCG, Collaboration, AI, Crowd Sourcing, Baba is You
\end{IEEEkeywords}

\IEEEpeerreviewmaketitle

\section{Introduction}

Level editors in games are relatively few in number but incredibly powerful. A game with a built-in level editor allows their players to continue the story and invites them to extend the game's system to new frontiers. As noted by Anna Anthropy, both hobbyist creators who are unfamiliar with coding and experienced coding wizards can use level editors to design new levels using game engines they are already familiar with \cite{anthropy2012rise}. Some games have level editors as part of a bonus feature for their games, such as Doom, the Tony Hawk series, and Halo 3, while other games are built with their entire gameplay focus on community submitted levels such as the Super Mario Maker series (Nintendo, 2015), Free Rider (Kano Games, 2006), and LittleBigPlanet (Media Molecule, 2008). With these tools, both the player and the game designers can explore what's possible given the constraints of the game engine's available mechanics. Not only this, but a creative process unfolds that can inspire future game designers to make levels with a diversity of mechanics used in interesting and original ways.

Artificially intelligent systems and agents have also been used in games to give unique experiences for each player. These systems typically work either during gameplay in the back-end as procedural content generators - creating new weapons (Borderlands 2), levels (Spelunky), or entire terrain environments (Minecraft). Artificial agents also help as collaborative companions in game to help the player complete the level - as either a puzzle solving partner (Shrek 2: The Video Game), an assistant fighter (the Pokemon Mystery Dungeon series), or simply as a guiding voice for the player (The Legend of Zelda: Ocarina of Time). However, there are very few AI systems or agents in games that collaborate with player creatively to help design levels. Mixed-initiative and collaborative PCG AI systems allow for a collaborative AI-user design loop to bridge this gap and allow both parties a chance to create content together, with the AI helping to guide the user towards specific design goals, adding new features, or testing for quality control. Both parties have their own design goals in mind, and both try to facilitate a two-way interaction in order to meet these goals and learn from each other in order to improve their designing abilities in ways that neither could achieve individually.

This paper presents a system that seeks to combine human design and AI-driven design to enable mixed-initiative collaborative game level creation. Users can choose to start from a blank slate with their work while adding their own edits then have an AI back-end evolve their work towards a pre-defined objective. This objective function can be defined by minimalism in design, maximization of game mechanic coverage, overall quality, or any other feature that could contribute to the quality of the level. Alternatively, users may select from a variety of AI suggestions and pre-generated samples to begin their work and then make changes as necessary. This design process is not limited to the initializing step of the level; the user and AI system can switch their roles as designers at any point in the creation process. Concurrently, the AI system will look at what its previous users have created and submitted, and ask new users to design levels that complement what's already there. With this design process, the mechanic space of a game can be fully explored and every combination of mechanics can be represented by a level. With a human-based rating system, the automated system can learn to design levels with better quality and the human users can design levels that are missing from this mechanic combination space. 

This project demonstrates the mixed-initiative collaborative process through level design for the independent, Sokoban-like game `Baba is You' - a game whose mechanics are defined and modified by the level design itself and the player's interaction with it. These ever changing mechanics expand the potential for levels with complex solutions and aesthetic designs and therefore offer a myriad of levels with a diverse array of mechanic combinations. Levels can be made either by users, AI, or a mixed combination of both and uploaded to the level database to be used for future creations and to improve the quality of the AI's objective function. The purpose of this system was to facilitate a collaborative interaction between an online community of level designers and PCG and recommendation AI system where both the user and the system can improve over time with increased interaction. Ideally, this could also encourage the development of more mixed-initiative HCI systems in the field of game design and the level editing games community as a whole.

This system was built on concepts from three different areas of content creation:
\begin{itemize}
    \item \textbf{Crowdsourcing:} a model used by different systems that allows a large set of users to contribute toward a common goal provided by the system~\cite{brabham2013crowdsourcing}. For example, Wikipedia users participate to fill in missing information for particular content.
    \item \textbf{User content creation:}
     allows players to create levels for a game/system and upload them online to the level database for other players to play and enjoy.
    \item \textbf{Quality diversity:} the underlying technique behind our system. It ensures that the levels made from combining the first 2 concepts are of both good quality and diverse in terms of the feature space they are established in~\cite{pugh2016quality}. For this system, the feature space is defined as the potential game mechanics implemented in each level.
\end{itemize}

\subsection{Baba is Y'all v1 (prototype)}

\begin{figure}[ht!]
    \centering
    \includegraphics[width=1.0\linewidth]{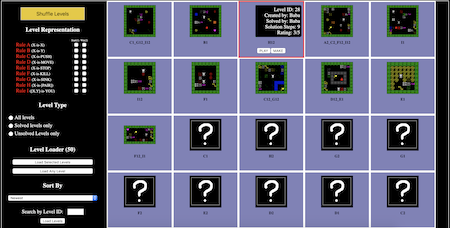}
    \caption{Baba is Y'all Version 1 Main Screen (from April 2020)}
    \label{fig:levelMat1}
\end{figure}
The first version of Baba is Y'all (BiY v1) was released officially in March 29th, 2020, and promoted chiefly on Twitter. This version served as a prototype and proof-of-concept system for mixed-initiative AI-assisted game content collaboration specifically for designing levels in the game `Baba is You' (Arvi 'Hempuli' Teikari, 2017).

The Baba is Y'all website (as shown in figure~\ref{fig:levelMat1}) was a prototype example of a mixed-initiative collaborative level designing system. However, the site was limited by the steep learning curve required to interact with the system~\cite{charity2020baba}. Features of the site were overwhelming to use and lack cohesion in navigating the site.

\subsection{Baba is Y'all v2 (updated release)}

\begin{figure}[ht!]
    \centering
    \includegraphics[width=1.0\linewidth]{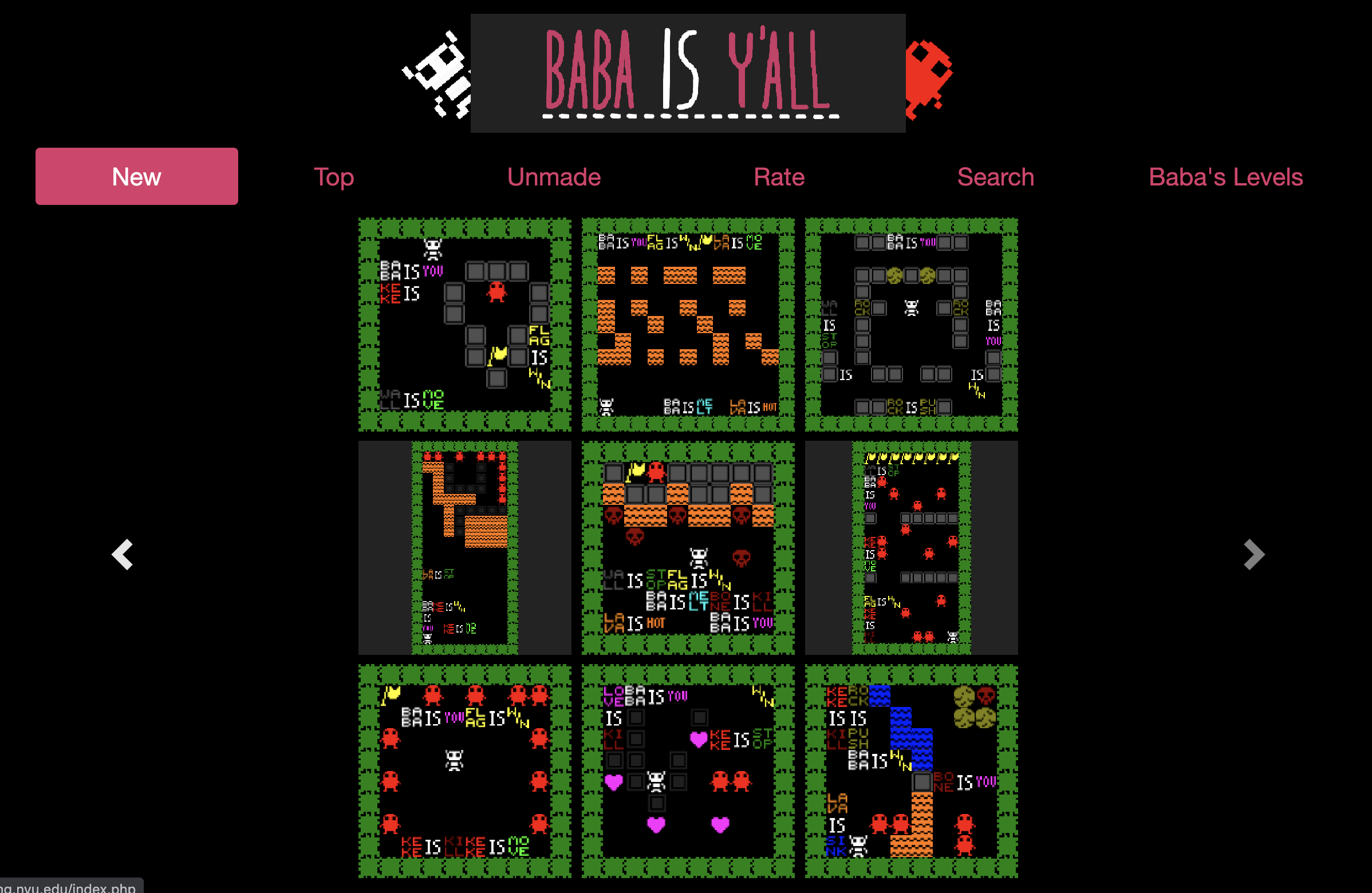}
    \caption{Baba is Y'all Version 2 Main Screen (as of September 2021)}
    \label{fig:levelMat2}
\end{figure}

The second version of Baba is Y'all\footnote{http://equius.gil.engineering.nyu.edu/} (BiY v2) was released on May 27th, 2021 and designed to have a more user-friendly setup. Figure~\ref{fig:levelMat2} shows the starting page of the home screen. It was similarly promoted via Twitter and on mailing lists. This version includes a cleaner, more compact, and more fluid user interface for the entire website and consolidated many of the separate features from the BiY v1 site onto fewer pages for easier access. Three main webpages were created for this updated system.


In addition to updating the features and collecting more data about the levels created, we conducted a formal user study with 76 participants to gather information about which features they chose to use for their level creation process and their subjective opinion on using the site overall. This user study, as well as the general level statistics collected from the site's database, showed that our new interface better facilitated the user-AI collaborative experience to create more diverse levels. We included this user study in order to better understand how users would approach our system and the artificially intelligent collaborative AI. With a formal survey and instructions for participants, we could examine how users choose to interact with the system to facilitate their creative goals and make inferences for how helpful the collaborative system actually is for the Baba is You level designing community.

\section{Background and Related Work}

The Baba is Y'all system uses the following methods in the collaborative level design process: procedural content generation to create new levels from the AI backend, quality diversity to maintain the different kinds of levels produced from the system and show the coverage of game mechanics across each level, crowdsourcing so the AI may learn to create new levels from previously submitted ``valid" levels - either those made exclusively by users, the system itself, or a combination of both, and finally mixed-initiative AI so that the user and evolutionary algorithm can develop the level together. Each method is described as the following: 

\subsection{Procedural Content Generation}
Procedural content generation (PCG) is defined as the process of using a computer program to create content that with limited or indirect user input \cite{shaker2016procedural}. Such methods can make an automated, quicker, and more efficient content creation process, and also enable aesthetics based on generation. PCG has been used in games from the 1980's Rogue to its descendent genre of the Rogue-likes used in games such as Spelunky (Mossmouth, LLC, 2008) and Hades (Supergiant Games, 2020), as well as games that revolve around level and world generation such as Minecraft (Mojang, 2011) and No Man's Sky (Hello Games, 2016). PCG can be used to build levels such as The Binding of Isaac (Edmund McMillen, 2011), enemy encounters such as Phoenix HD (Firi Games, 2011), or item or weapon generation such as Borderlands (Gearbox Software, 2009). In academia, PCG has been explored in many different game facets for generating assets \cite{ruela2017procedural, gonzalez2020generating}, mechanics \cite{khalifa2019general, browne2010evolutionary}, levels \cite{snodgrass2016learning,charity2020mech}, boss fights~\cite{siu2016programming}, tutorials~\cite{khalifa2019intentional,green2018atdelfi}, or even other generators \cite{kerssemakers2012procedural,earle2021illuminating}. 

A plethora of AI methods underpin successful PCG approaches, including evolutionary search \cite{togelius2010search}, supervised and unsupervised learning \cite{summerville2018procedural,liu2021deep}, and reinforcement learning \cite{khalifa2020pcgrl}. The results of these implementations have led to PCG processes being able to generate higher quality, more generalizable, and more diverse content. PCG is used in the Baba is Y'all system to allow the mutator module to create new `Baba is You' levels.

\subsection{Quality Diversity}
Quality-diversity (QD) search based methods are increasing in usage for both game researchers and AI researchers \cite{pugh2016quality,gravina2019procedural}. Quality-diversity techniques are search based techniques that try to generate a set of diverse solutions while maintaining high level of quality for each solution. A well-known and popular example is MAP-Elites, an evolutionary algorithm that uses a multi-dimensional map instead of a population to store its solutions~\cite{mouret2015illuminating}. This map is constructed by dividing the solution space into a group of cells based on a pre-defined behavior characteristics. Any new solution found will not only be evaluated for fitness but also for its defined characteristics then placed in the correct cell in the MAP-Elites map. If the cell is not empty, both solutions compete and only the fitter solution survives. Because of the map maintenance and the cell competition, MAP-Elites can guarantee a map of diverse and high quality solutions, after a finite number of iterations through the generated population. For this project, we use the Constrained MAP-Elites algorithm \cite{khalifa2019intentional, alvarez2019empowering} to maintain a diverse population of `Baba is You' levels where the behavior characteristic space of the matrix is defined by the starting and ending rules of a level when it is submitted. 

\subsection{Crowdsourcing data and content}
Some, but relatively few, games allow users to submit their own custom creations using the game's engine as most games do not have their source code available or even partially accessible for modifications to add more content in the context of the game. Whether through a built-in level editing system seen in games like Super Mario Maker (Nintendo, 2015), LittleBigPlanet (MediaMolecule, 2008), or LineRider (inXile Entertainment, 2006) or through a modding community that alter the source code for notable games such as Skyrim (Bethesda, 2011) Minecraft (Mojang, 2011,) or Friday Night Funkin' (Ninjamuffin99, 2020), players can create their own content to enhance their experience and/or share with others. While a modding system is an indirect and unfacilitated product of a game, both the modding community and the level editor community look to improve or expand on the base gameplay to create unique and interesting experiences for players while keeping to the original domain that the original designers may not have foreseen.

In crowdsourcing, many users contribute data that can be used for a common goal. Some systems like Wikipedia rely entirely on content submitted by their user base in order to provide information to others on a given subject. Other systems like Amazon's MechanicalTurk crowdsource data collection, such as research experiments \cite{buhrmester2016amazon}, by outsourcing small tasks to multiple users for a small wage. An example of a game generator based on crowdsourced data is Barros et al.’s DATA Agent \cite{barros2018killed}, which uses crowd-sourced data such as Wikipedia to create a point-click adventure game sourced from a large corpus of open data to generate interesting adventure games.

What differentiates the Baba is Y'all system from other level editing systems or interactive PCG systems is that the Baba is Y'all site has a central goal: populate the MAP-Elites matrix with levels that cover all possible rule combinations. With this system, users may freely create the levels they want, but they may also work towards completing the global goal of making levels with a behavior characteristic that has not been made before. Participation in this task is encouraged by the AI back-end system that keeps track of missing cells in the MAP-Elites matrix.

\subsection{Mixed-Initiative AI}
Mixed-initiative AI systems involve a co-creation of content between a human user and an artificially intelligent system~\cite{yannakakis2014mixed}. Previous mixed-initiative systems include selecting from and evolving a population of generated images \cite{secretan2008picbreeder}, composing music \cite{mann2016ai,tokui2000music}, and creating game levels through suggestive feedback \cite{machado2019pitako}. Mixed-initiative and collaborative AI level editors for game systems have thoroughly been explored in the field as well through direct and indirect interaction with the AI backend system \cite{shaker2013ropossum,liapis2013sentient,butler2013mixed,guzdial2018co,zhou2021toward,bhaumik2021lode,smith2010tanagra,delarosa2021mixed}. Compton's definition of casual creation tools and survey of existing systems also strongly fit for the objective of mixed-initiative systems, where the end goal may not just involve creating a productive output, but facilitating a creative experience \cite{compton2015casual}.

Since the release of the first Baba is Y'all prototype and paper~\cite{charity2020baba}, the implementation of mixed-initiative systems have grown in the game and AI research field. Bhaumik implemented an AI constrained system with their Lode Encoder level editing tool that only allowed users to edit a level from a set of levels generated by a variational autoencoder - forcing users to only edit from a palette provided by the AI back-end tool~\cite{bhaumik2021lode}. Delarosa used a reinforcement learning agent in a mixed-initiative web app to collaboratively suggest edits to Sokoban levels \cite{delarosa2021mixed}. Zhou used levels generated with the AI-assisted level editor Morai Maker (a Super Mario level editor) to apply transfer learning for level editing to Zelda \cite{zhou2021toward}. These recent developments look more into how the human users are affected through their relationship with collaborating with these AI systems and how it can be improved through examining the dimensionality of the QD algorithm, the evolutionary process, or the human-system interaction itself \cite{alvarez2020exploring}. We look to incorporate these new perspectives into this updated iteration of Baba is Y'all and evaluate the effects through a user study.

\section{System Description}

\begin{figure}[ht!]
    \centering
    \includegraphics[width=1.0\linewidth]{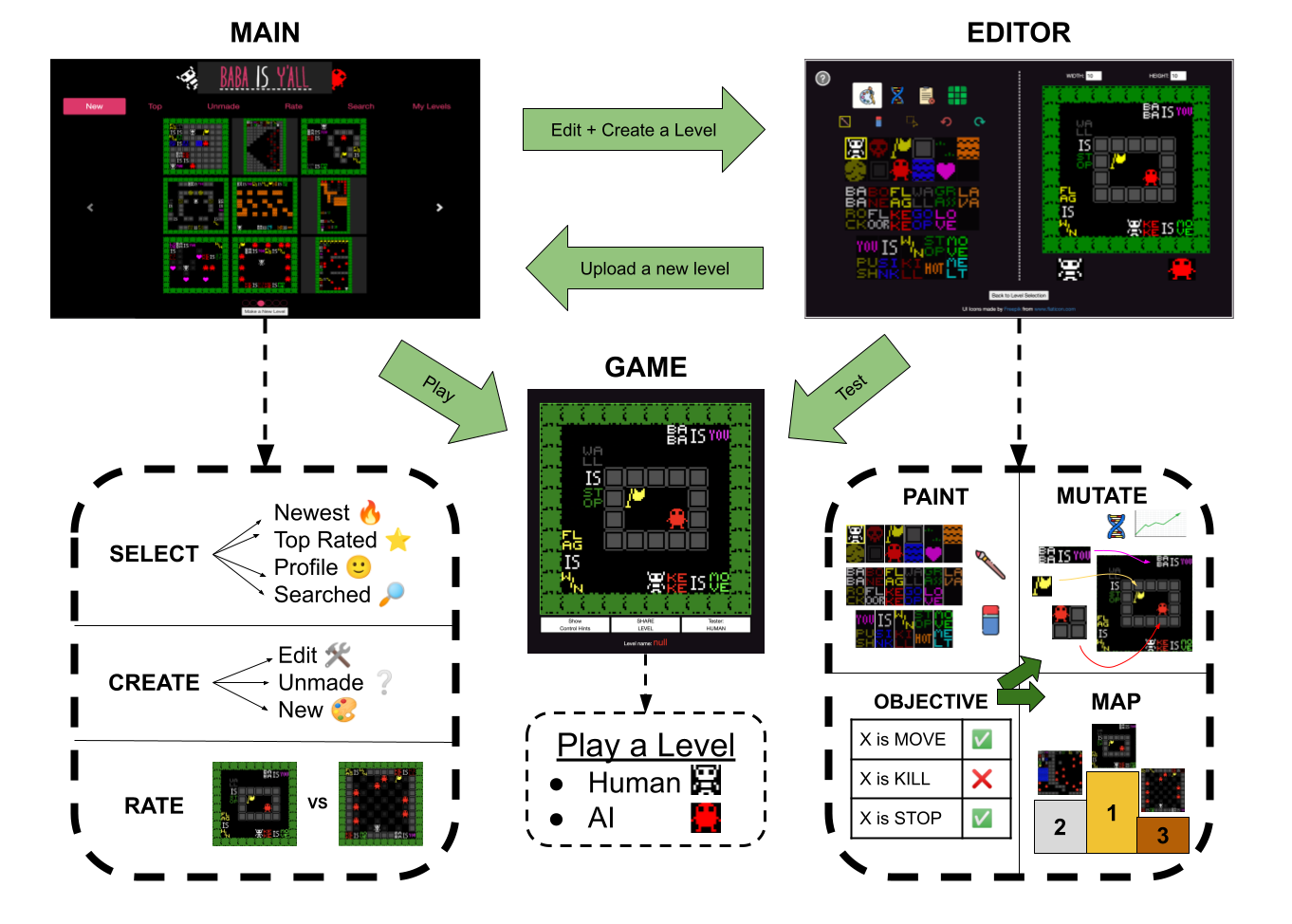}
    \caption{A system's diagram of the updated Baba is Y'all website.}
    \label{fig:system_flowchart}
\end{figure}

While acting as a great prototype, the first version of the Baba is Y'all site was arguably overwhelming to use - as it showed most of the features and data all at once and required the user to navigate through various pages and inputs to use the system. There was also a lack of documentation or tutorialization to help users acclimate to the site. Users had to switch between two pages in order to manually edit their levels and to have the back-end AI system evolve their levels. The levels shown on the main screen were also unorganized and randomly shown to the user, with the published levels and unmade rule combination levels added together on the same visual grid. We developed a new Baba is Y'all interface and system shown in a system diagram in Figure \ref{fig:system_flowchart}. The updated Baba is Y'all site's features were condensed into 2 main pages to make navigation and level editing much easier and intuitive:
\begin{itemize}
    \item \textbf{The Home Screen:} contains the level matrix \textit{Map Module}, the search page, the \textit{Rating Module} page, and the \textit{User Profile} page. From here, users can also change the visuals of the site from light to dark mode, view the tutorial section or the site stats page by clicking on the Baba and Keke sprites respectively at the top of the page, and create a new level from scratch by clicking on various 'Create New Level' buttons placed on various subpages.
    \item \textbf{The Level Editor Screen:} contains both the \textit{Editor Module} and the \textit{Mutator Module}. Users can also test their levels with themselves or with the Keke solver by clicking on the Baba and Keke icons at the bottom of the canvas. Figure~\ref{fig:editor_screen} shows the starting page of the level editor screen.
\end{itemize}

Unlike the previous version, which showed all of the mechanic combination levels (both from the database and unmade) in random order, the updated level selection page adds level tabs that separates levels by recently added (New), highest rated (Top), and levels with rules that had not been made yet (Unmade.) A carousel scrolling feature shows 9 levels at a time to not overwhelm the player with choices (as shown in figure~\ref{fig:levelMat2}). The level rating system is also included on the main page as a tab, as well as the search feature. The personal level selection tab allows users to see their previously submitted levels and login to their account to submit levels with their username as the author or co-author.

The updated level editing page consolidates both the user editing with the PCG level evolution onto one page. Users can easily switch between manually editing the level themselves and allowing the PCG back-end system to edit the level while pausing in between. Users can also select rule objectives for the system to evolve towards implementing. To fight the problem of blank canvas paralysis, users can start from a set of different types of levels (both PCG and user-made)~\cite{krall2012artist}. Once a level is successfully solved, users may name the level upon submission - further personalizing the levels and assigning authorship.

A slideshow tutorial is provided for the users and describes every feature and function of the site instead of the walkthrough video that was featured on BiY v1. Users can also play a demo version of the `Baba is You' (Arvi 'Hempuli' Teikari, 2017) game to familiarize themselves with the game mechanics/rule space and how they interact with each other (game dynamics). For quick assistance, a helper tool is provided on the level editing page as a refresher on how to use the editing tool.

In the following subsections, we are going to explain the different modules that constitutes these two main screens. Each of the following modules are either being used in the home screen, the level editor screen, or both.

\subsection{Baba is You}

`Baba is You' (Arvi ``Hempuli'' Teikari, 2019) is a puzzle game where players can manipulate the rules of a level and properties of the game objects through Sokoban-like movements of pushing word blocks found on the map. These dynamically changing rules create interesting exploration spaces for both procedurally generating the levels and solving them - thus making it a viably complex domain for a mixed-initiative level editor system. The different combinations of rules can also lead to a large diversity of level types that can be made in this space. 


The general rules for the `Baba is You' game can be referred to from our previous paper~\cite{charity2020baba}. To reiterate, there are three types of rule formats in the game:
\begin{itemize}
    \item \textbf{X-IS-(KEYWORD)} a property rule stating that the game object class `X' has a certain property such as `WIN', `YOU', `MOVE', etc.
    \item \textbf{X-IS-X} a reflexive rule stating that the game object class `X' cannot be changed to another game object class.
    \item \textbf{X-IS-Y} a transformative rule changing all game objects of class `X' into game objects of class `Y'.
\end{itemize}



\begin{figure}[ht!]
    \centering
    \includegraphics[width=1.0\linewidth]{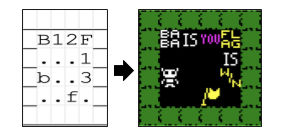}
    \caption{Example of an ascii representation of a level to a render of the level.} 
    \label{fig:ascii_map}
\end{figure}

The game sprites are divided into two main different classes: the object class and the keyword class. Sprites in the object class represent the interactable objects in the map as well as the literal word representation for the object. Sprites in the keyword class represent the rules of the level that manipulate the properties of the objects. For example, figure~\ref{fig:ascii_map} shows four different object class sprites [BABA (object and corresponding word) and FLAG (object and corresponding word)] and three different keyword class sprites [IS (x2), YOU, and WIN]. The keyword class sprites are arranged in two rules: `BABA-IS-YOU' allowing the player to control all the Baba objects and `FLAG-IS-WIN' indicating that reaching any flag object will make the player win the level. The system has a total of 32 different sprites: 11 object class sprites and 21 keyword class sprites. Because the game allows rule manipulation, object classes are arbitrary in the game as they serve only to provide a variety of objects for rules to affect and for aesthetic pleasure. 


\subsection{Game Module}

The game module is responsible for simulating a `Baba is You' level. It also allows users to test the playability of levels either by directly playing through the level themselves or by allowing a solver agent to attempt to solve it. This component is used on the home screen when a user selects a level to play and the editor screen for a user to test their created level.

Because the game rules are dynamic and can be altered by the player at any stage in the solution, the system keeps track of all the active rules at every state. Once the win condition has been met, the game module records the current solution, the active rules at the start of the level, and the active rules when the solution has been reached. These properties are saved to be used and interpreted by the Map module (section~\ref{sec:map_module}). 
The activated rules are used as the level's characteristic feature representation and saved as a chromosome to the MAP-Elites matrix.

The game module provides an AI solver called 'KEKE' (based on one of the characters traditionally used as an autonomous 'NPC' in the game). This solver was provided to give users the choice to quickly test their levels for solvability and for complexity of solutions. KEKE uses a greedy best-first tree search algorithm that tries to solve the input level. The branching space is based on the five possible inputs a player can do within the game: move left, move right, move up, move down, and do nothing. The algorithm uses a heuristic function based on a weighted average of the Manhattan distance to the centroid distance for 3 different groups: keyword objects, objects associated with the `WIN' rule, and objects associated with the `PUSH' rule. These were chosen based on their critical importance for the user solving the level - as winning objects are required to complete the level, keyword objects allow for manipulation of active rules, and pushable objects can directly and indirectly affect the layout of a level map and therefore the accessibility of player objects to reach winning objects. The heuristic function is represented by the following equation:
\begin{equation}
    h = (n + w + p) / 3
\end{equation}
where $h$ is the final heuristic value for placement in the priority queue, $n$ is the minimum Manhatttan distance from any player object to the nearest winnable object, $w$ is the minimum Manhatttan distance from any player object to the nearest word sprite, and $p$ is the minimum Manhatttan distance from any player object to the nearest pushable object.

As an update for this version of the system, the agent can run for a maximum of 10000 iterations (10 times more iterations than previous the previous solver to allow for more searching) and can be stopped at any time by the user. After stopping, a user may also attempt to solve part of the level themselves and the KEKE solver can pick up where the user left off to attempt to solve the remainder of the level. This creates a mixed-initiative approach to solving the levels in addition to editing the levels. However, even with this collaborative approach, the system still has limitations and difficulty solving levels with complex solutions - specifically solutions that require back-tracking across the level after a rule has been changed. The solver runs on the client side of the site and is limited by the capacity of the user's computational resources. Future work will look into improving the solver system to reduce computational resource. We will also look for better solving algorithms to improve the utility of the solver such as Monte Carlo Tree Search (MCTS) with reversibility compression~\cite{cook2021monte}.

\subsection{Editor Module}

\begin{figure}[ht!]
    \centering
    \includegraphics[width=1.0\linewidth]{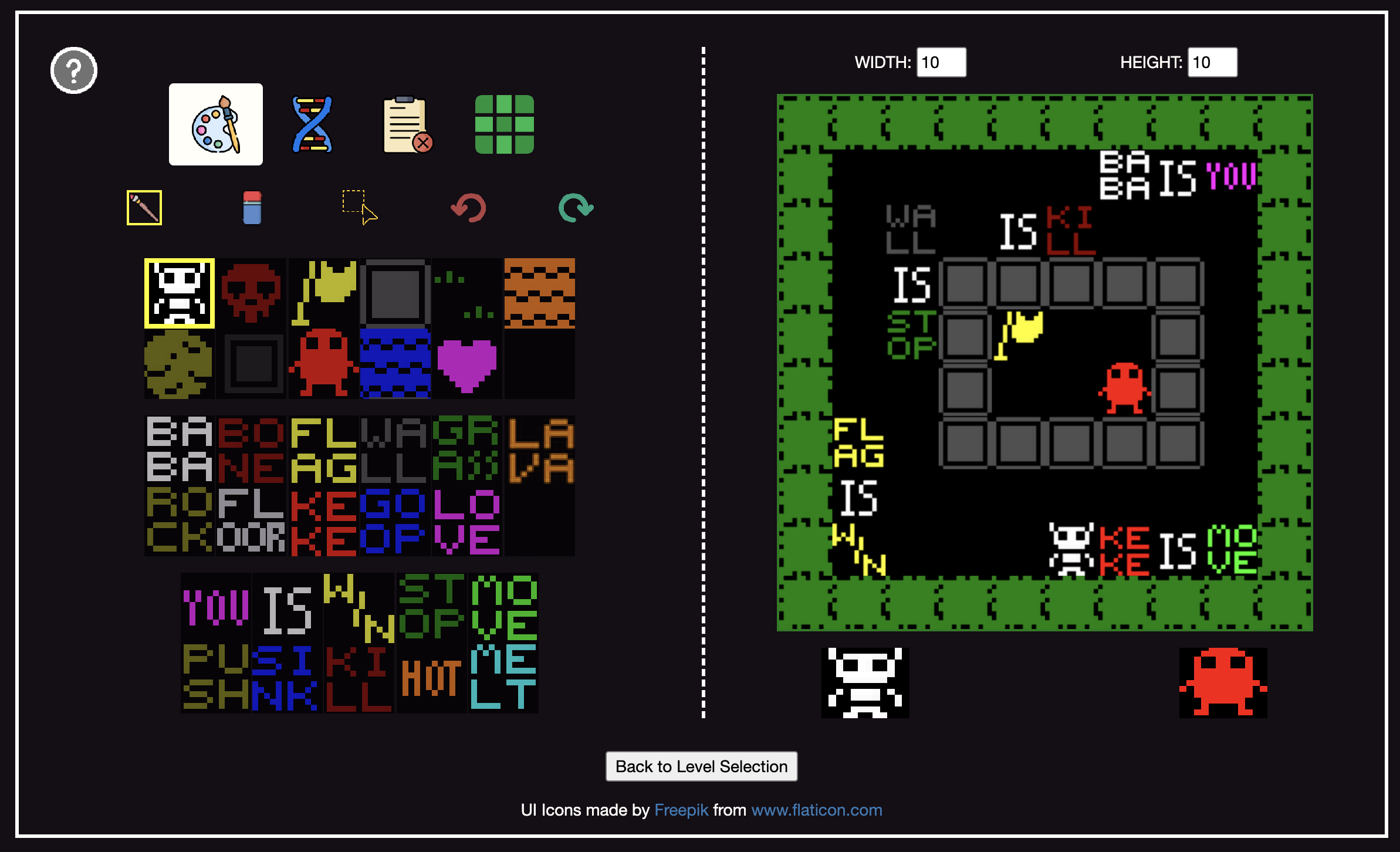}
    \caption{A screenshot of the level editor screen} 
    \label{fig:editor_screen}
\end{figure}

The editor module of the system allows human users to create their own `Baba is You' levels in the same vain of Super Mario Maker (Nintendo, 2015). Figure~\ref{fig:editor_screen} shows the editor window that is available for the user. The user can place and erase any game sprite or keyword at any location on the map using the provided tools. As a basis, the user can start modifying either a blank map, a basic map (a map with X-IS-YOU and Y-IS-WIN rules already placed with X and Y objects), a randomly generated map, or an elite level provided by the Map Module. Similar to Super Mario Maker (Nintendo, 2015), the created levels can only be submitted after they are tested by the human player or the AI agent to check for solvability. For testing the level, the editor module sends the level information to the game module to allow the user to test it. 

This updated version of the site also includes an undo and redo feature so that users may erase any changes they make. A selection and lasso feature is also available so users can select specific areas of the level and move them to another location. Unlike the previous version, all tiles are available to the user on the same screen and the user may seamlessly transition from the editor module to the mutator module and vice versa for ease of access and better interactivity and collaboration between the AI system and the user. We included all of the tiles on the same screen along with the new tools to allow for a more seamless design process and to make the manual editing more intuitive for users - with the interface design intended to resemble familiar computer art programs such as MSPaint, GIMP, PaintTool SAI, or Photoshop.

\subsection{Mutator Module}\label{sec:mutator_module}

\begin{figure}[ht!]
    \centering
    \includegraphics[width=1.0\linewidth]{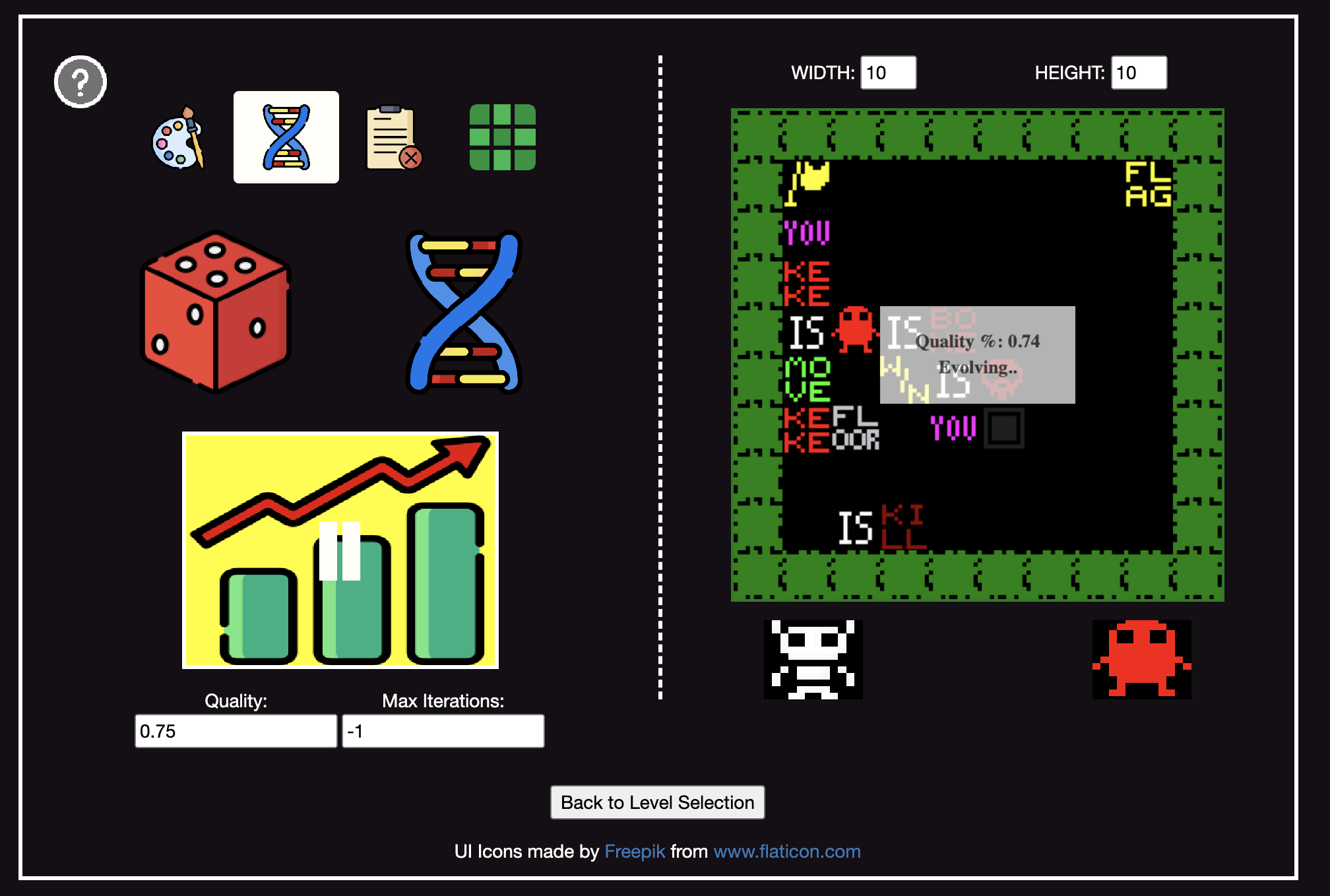}
    \caption{A screenshot of the level evolver page}
    \label{fig:evolver_screen}
\end{figure}

The Mutator module is a procedural content level generator. More specifically, the Baba is Y'all system uses an evolutionary level generator that defines a fitness function based on a version of tile-pattern Kullback-Liebler Divergence (ETPKLDiv\footnote{https://github.com/amidos2006/ETPKLDiv}) algorithm~\cite{lucas2019tile}. The mutator module creates an ``algorithm-centric'' approach to creating levels, where the levels are generated towards a fitness value and can then edited by the user later. Within the level map population, the mutator modifies the ascii-representation of the level itself by replacing sprites and tiles. Figure~\ref{fig:evolver_screen} shows the updated interface used by the evolver. As mentioned before in the previous subsection, this version of the mutator module can interface seamlessly with the other modules to allow the user more ease of access between manual editing and evolutionary editing. The user can easily transfer the level from the editor module to the mutator module and vice versa. 
When switching between the editor module and the mutator module, the level loses its pure procedurally generated or pure human-designed quality and becomes a hybrid of the two - thus mixed-initiative interaction between the algorithm and the user.

The interface screen provides the user with multiple customizations such as the initialization method, stopping criteria, evolution pausing, and an application of a mutation function allowing manual user control. With these features, the user is not directly changing the evolution process itself, but instead guiding and limiting the algorithm towards generating the level they want. We designed this interface with the intention to make the evolutionary process as transparent as possible for more advanced users so that they can adjust the mutator's editing abilities to their own criteria and standards while keeping it simple and general for more intermediate users who were unfamiliar with PCG or collaborative design systems.

The ETPKLDiv algorithm uses a 1+1 evolution strategy, also known as a hillclimber, to improve the similarity between the current evolved levels and a reference level. The algorithm uses a sliding window of a fixed size to calculate the probability of each tile configuration (called tile patterns) in both the reference level and the evolved level and tries to minimize the Kullback-Liebler Divergence between both probability distributions.

Like Lucas and Volz, we use a window size of 3x3 for the tile selection. This was to maximize the probability of generating initial rules for a level, since rules in `Baba is You' are made up of 3 tiles. However, in our project, we used 2+2 evolution strategy instead of 1+1 used to allow slightly more diversity in the population~\cite{lucas2019tile}. We also modified the fitness function to allow it to compare with more than one level. The fitness value also includes the potential solvability of the level ($p$), the ratio of empty tiles ($s$), and the ratio of useless sprites ($u$).  
The final fitness equation for a level is as follows:
\begin{equation}\label{eq:fitness}
    fitness_{new} = min(fitness_{old}) + u + p + 0.1 \cdot s
\end{equation}
where $fitness_{old}$ is the Kullback-Lievler Divergence fitness function from the Lucas and Volz work~\cite{lucas2019tile} compared to a reference level. The minimum operator is added as we are using multiple reference levels instead of one and we want to pick the fitness of the most similar reference level.

In the updated version of Baba is Y'all, we recalculate the ratio of useless objects ($u$) used in the original version's equation. The value $u$ is defined as the combined percentage of unnecessary object and word sprites in the level. This is broken up into 2 variables $o$ and $w$ for the objects and words respectively. The $o$ value corresponds to the ratio of objects in the initial state of the level that are not required or predicted to act as a constraint or solution. The value for $o$ can be calculated as follows:
\begin{equation}
    o = \frac{i}{j}
\end{equation}
where $i$ is the number of objects sprites initialized in the level without a related object-word sprite and $j$ is the total number of object sprites initialized in the level. While the $w$ value corresponds to the ratio of word sprites that have no associated object in the map to all of the word sprites in level (this does not apply to keyword class words such as ``KILL'' or ``MOVE''.) The value for $w$ can be calculated as follows: 
\begin{equation}
    w = \frac{k}{l}
\end{equation} 
where $k$ is the number of word sprites initialized in the level without a related object-word sprite and $l$ is the total number of word sprites initialized in the level. To combine both variables $o$ and $w$ into the one variable $u$ a constant ratio is applied. In the system, 0.85 is applied to the $o$ variable and 0.15 to $w$. This is to more weight on reducing the number of useless object sprites as opposed to useless word sprites, as word sprites can be used to modify the properties of objects or transform other object sprites.

The $u$ value is implemented in order to prevent noise within the level due to having object tiles that cannot be manipulated in any way or have relevancy to the level. A human-made level may include these ``useless'' tiles for aesthetic purposes or to give the level a theme - similar to the original `Baba is You' levels. However, the PCG algorithm optimizes towards efficiency and minimalist levels, therefore ignoring the subjective aspect of a level's quality (which can be added later by the user).

The playability of the level ($p$) is a binary constraint value that determines whether a level is potentially winnable or not. The value can be calculated as follows:
\begin{equation}
    p = 
    \begin{cases}
        0, & \text{has [`X-IS-YOU' rule, `WIN' keyword]} \\
        1, & otherwise
    \end{cases}
\end{equation}
This is to ensure any levels that are absolutely impossible to play or win are penalized in the population and less likely to be mutated and evolved from in future generations. We used a simple playability constraint check instead of checking for playability using the solver because the solver take time to check for playability. Also, all playable levels by the solver usually end up being easy levels due to the limited search space we are given for the best first algorithm.

The ratio of empty tiles ($s$) is the ratio of empty space tiles to all of the tiles in the level. The equation can be calculated as follows:
\begin{equation}
    s = \frac{e}{t}
\end{equation}
where $e$ is the number of empty spaces in the level and $t$ is the total number of tiles found in the level. The value $s$ is multiplied with a value of $0.1$ in equation~\ref{eq:fitness} to avoid heavy penalization for having any empty spaces in a level and to prevent encouragement for levels to mutate towards populating the level with an overabundance of similar tiles in order to eliminate any empty space.

\begin{figure}[ht!]
    \centering
    \includegraphics[width=1.0\linewidth]{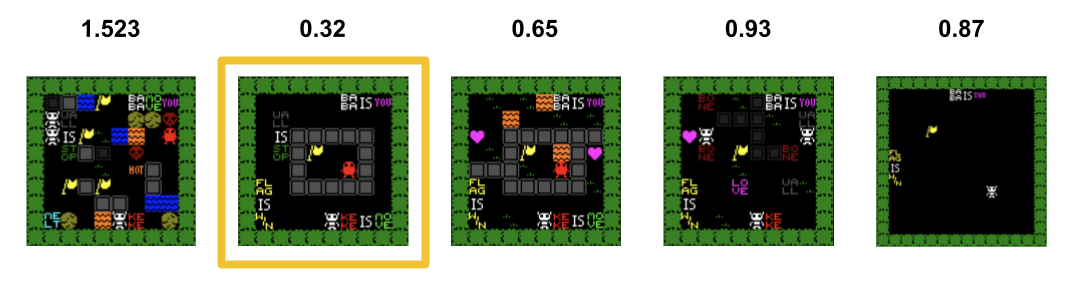}
    \caption{A sample population selecting a map with the best fitness value from after multiple iterations of evolution.} 
    \label{fig:mut_examp}
\end{figure}

Figure \ref{fig:mut_examp} demonstrates an example of the selection process from the evolutionary algorithm's population using the fitness function defined. For this example, the selected objective rules the user wants the level to evolve towards is to contain or potentially contain the rules 'X-is-STOP' and 'X-is-MOVE.' The values at the top of the maps represent each map's fitness value. The first map from the left has a bad fitness value from having too many irrelevant sprites, not fitting the objective criteria, and is definitively unsolvable, since no 'WIN' sprite tile or 'YOU' sprite tile exists. The middle map is substantially better and fits the objective criteria, but has many unreferenced tiles (as no word sprite block for 'GRASS', 'LAVA', or 'LOVE' exists on the map.)  The evolver is intended to maximize towards minimalism, so it would likely try to remove these sprites. However, these tiles could arguably serve as decoration instead and make the level more aesthetically pleasing and a user could edit them manually back into the level if they wanted. The two levels on the far right are also lower in quality and fitness from not including the word blocks necessary to match the objective request and from having too much empty space. Therefore, the second map from the left will be selected to be shown to the user during the evolutionary process and selected as the most viable map for the user. 

The Mutator module is not run as a back-end process to find more levels, instead it has to be done manually by the user. This is done due to the fact that some generated levels cannot be solved without human input. One might wonder why not generate a huge corpus of levels and ask the users later to test them for the system. This could result in the system generating a multitude of levels that are either impossible to solve or are solvable but not subjectively ``good'' levels - levels the user would not find pleasing or enjoyable. This overabundance of ``garbage'' levels could lead to a waste of memory and a waste human resources. By allowing the user direct control over which levels are submitted from the generation algorithm, it still guarantees that the levels are solvable and with sufficient quality and promote using the tool in a mixed-initiative approach. Future work will explore implementing a fully autonomous generator and associated solver to expand the archive of levels without human input.

\subsection{Objective Module}\label{sec:objective_module}

\begin{figure}[ht!]
    \centering
    \includegraphics[width=1.0\linewidth]{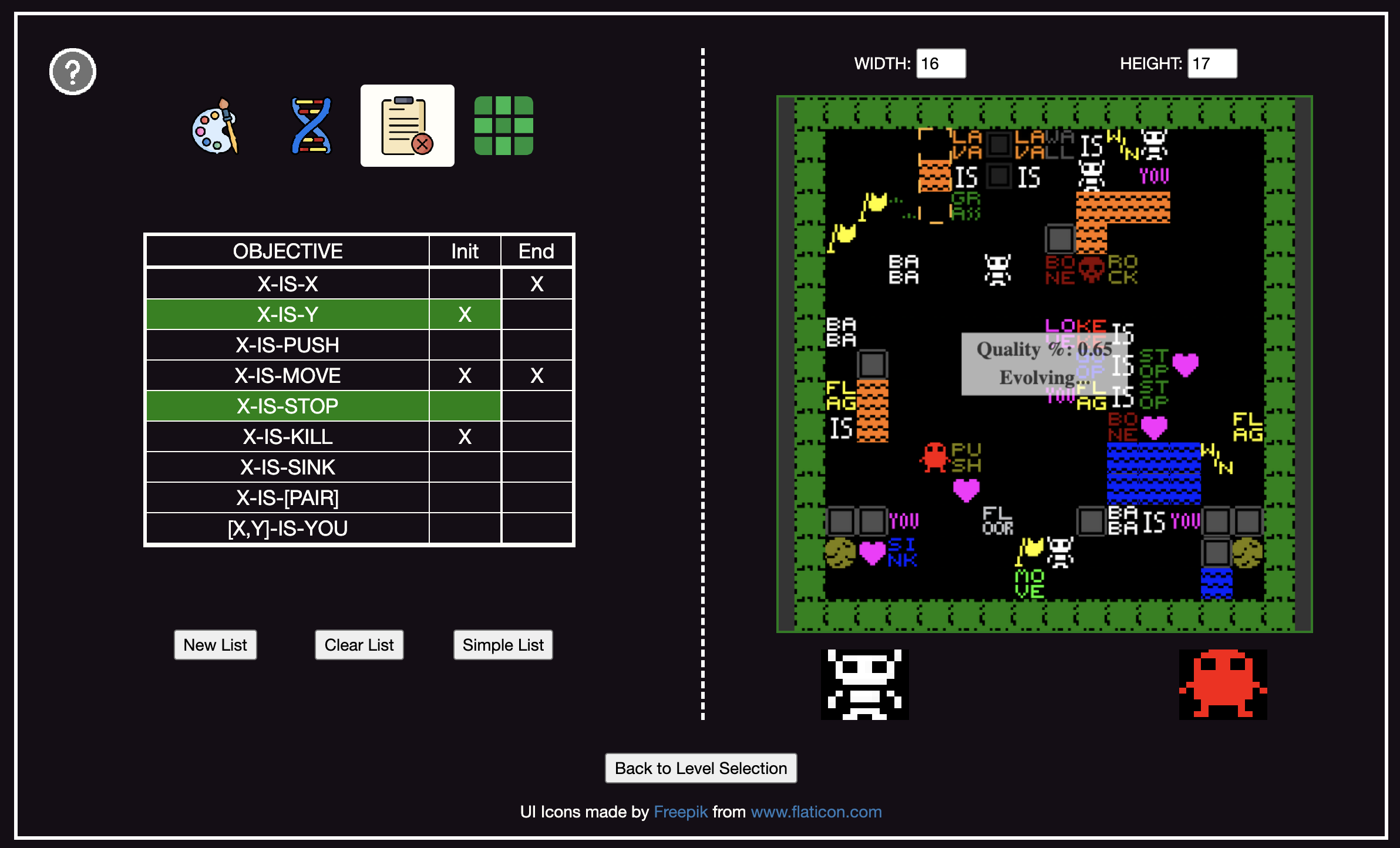}
    \caption{A screenshot of the rule objective screen} 
    \label{fig:objective_screen}
\end{figure}

In conjunction with the Mutator module (section~\ref{sec:mutator_module}), an Objective Module has been implemented to help guide the evolver towards generating levels that match selected objectives - or rules - set by either the Map Module or the user. Like the original Baba is Y'all website, each level has can have a set of starting and ending rules - the objective module allows users to select which rules they would like to be present (or potentially be present) in the level. The evolver module will then use this rule selection in order to optimize evolving maps that could accomplish the rule objectives by adding more word sprites or combined word groups related to the objectives that a player could either have at the beginning of the level or use for the solution of the level. Like before this will nudge both the user and the evolver back-end towards creating levels with mechanic combinations that have not been made in the site database. This module was implemented with the AI back-end to place more focus on designing diverse levels that have not previously been made or saved to the site database and therefore expand the mechanic space of the levels overall. 

Users can select from the table of mechanics (shown in Figure \ref{fig:objective_screen}) which sets of rules to include in the level - whether initially at the start of the level, at the solution, or either. Initial rules can be found automatically when the user or evolver edits the level, final rules can only be determined at the end of the level - when the solution has been found. Active rules are highlighted with a green backlight in the table and change accordingly when a rule is created or removed. 

The evolver also prioritizes levels that match as many of the selected rules as possible. A cascading function is used to rank the generated levels from the chromosome population. The evolver first evaluates how well a generated level corresponds to the selected objectives then looks at the fitness function. With this, the evolver becomes more involved with expanding the level database for the site and actively tries to help the user fill these missing levels.

\subsection{Rating Module}\label{sec:rating_module}

\begin{figure}[ht!]
    \centering
    \includegraphics[width=1.0\linewidth]{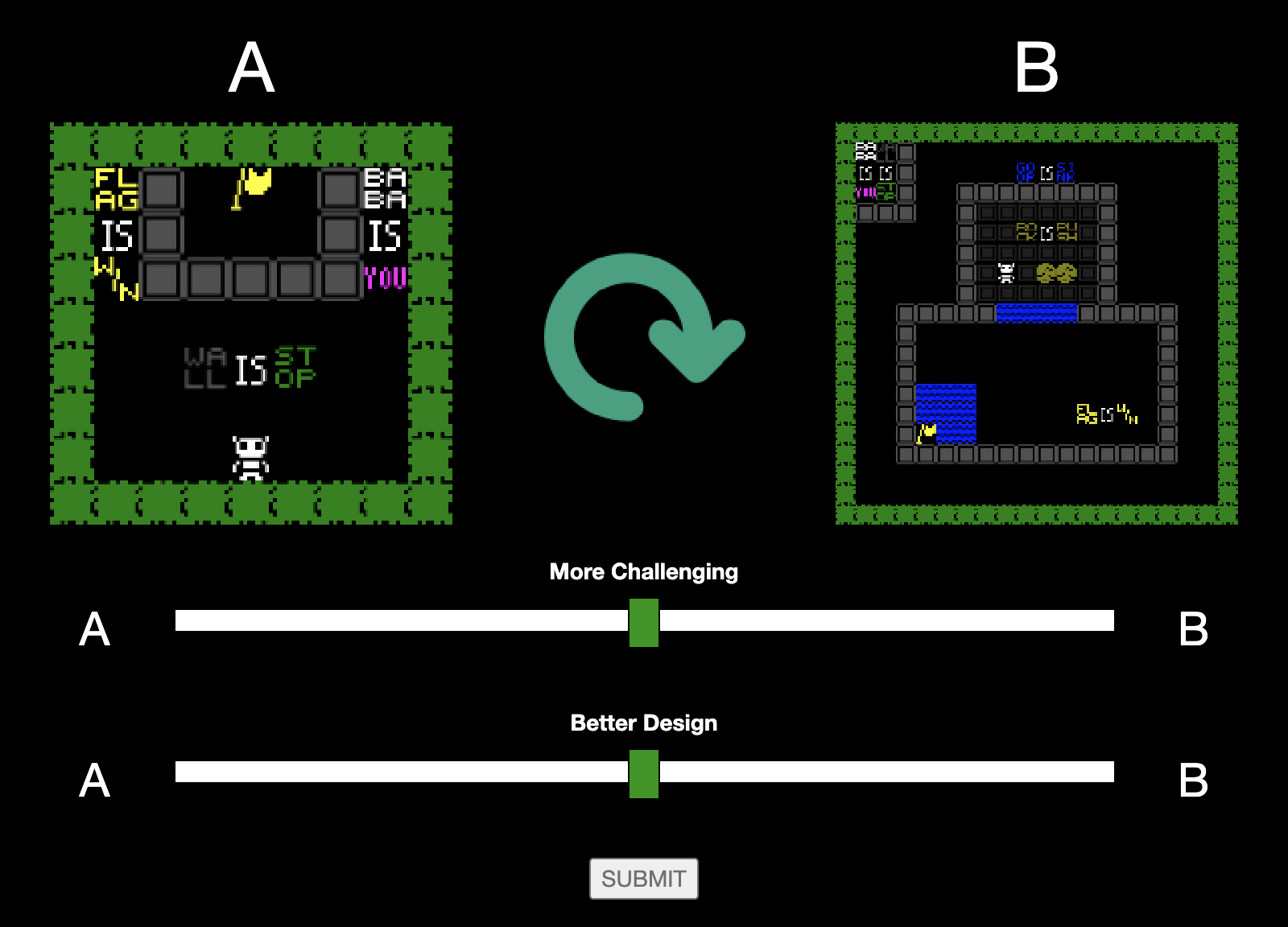}
    \caption{A screenshot of the rating screen with 2 levels shown} 
    \label{fig:rating_screen}
\end{figure}

Like the original system, a rating for a single level is determined by comparison to another level within the site database. The user must determine the better level based on two qualities: level of challenge and quality of aesthetic design. A level that is considered `more challenging' could indicate that the solution search space for the level takes longer to arrive at or is not as intuitive or straightforward. A level that is considered to have `better design' represents that the level is more visually pleasing and elegant with its map representation - a quality that is hard to generate automatically with AI. Users can select between the two levels for each feature by shifting a slider towards one level or the other. The rating system is implemented to teach the AI back-end system which levels have higher ``quality" and to use within the ranking and recommendation of the editor and mutator modules. Figure \ref{fig:rating_screen} shows a screenshot of the rating page on the site that allows user to change the sliding bar value between two compared levels to evaluate for challenge level and aesthetic design.

\subsection{Map Module}\label{sec:map_module}

\begin{figure}[ht!]
    \centering
    \includegraphics[width=1.0\linewidth]{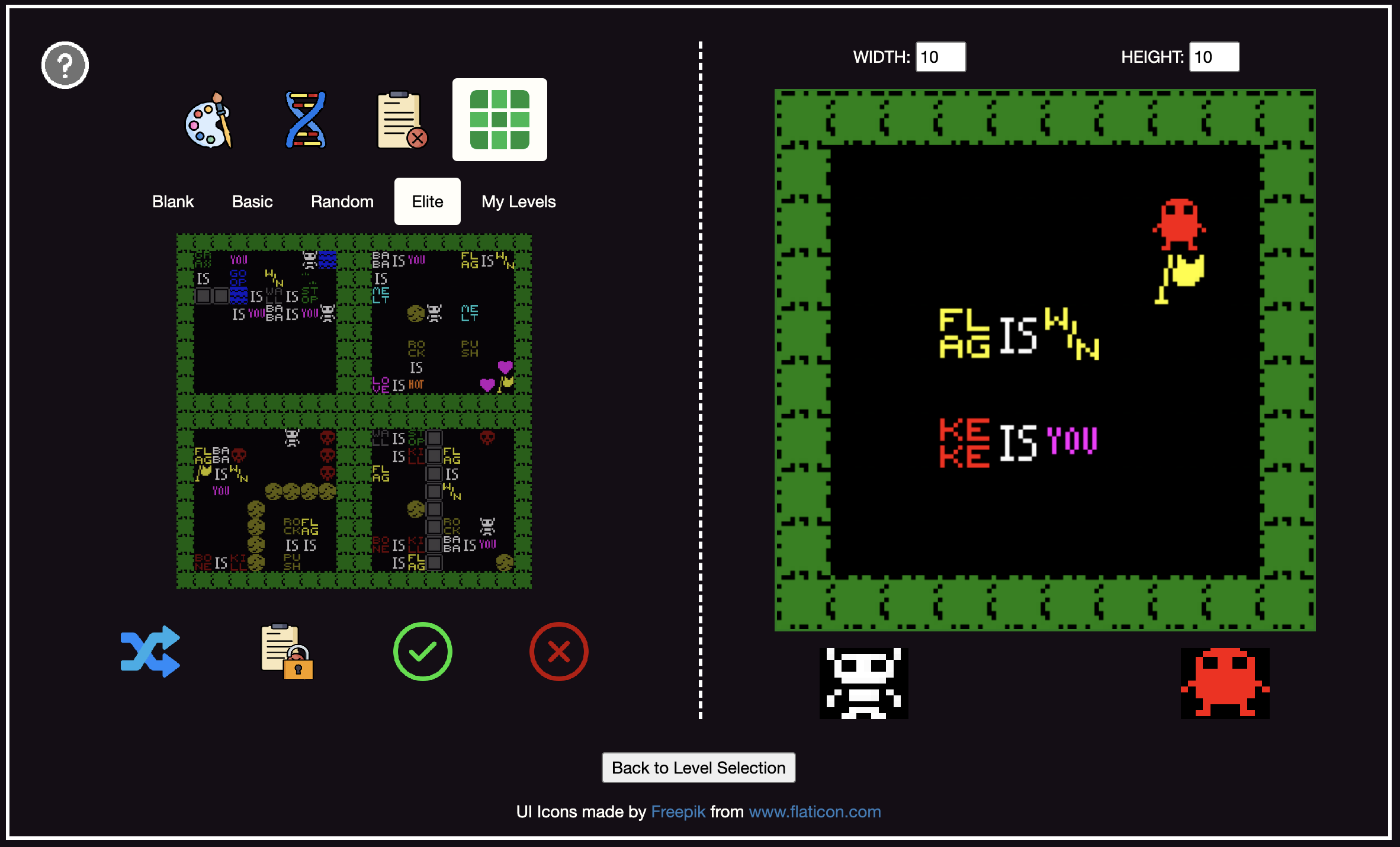}
    \caption{A screenshot of the map selection screen} 
    \label{fig:select_screen}
\end{figure}

The Map module functions as both storing all of the levels in the site database as well as recommending specific levels to the user to use for their own level creation process. The Map module is the core module of the system. To maintain distinguish-ability between quality and diverse levels, we implemented the MAP-Elites algorithm for this module.

\begin{table}[t]
    \caption{Chromosome Rule Representation}
    \centering
    \begin{tabular}{|p{0.2\linewidth}|p{0.7\linewidth}|}
    \hline
         Rule Type & Definition \\
    \hline
    \hline
        X-IS-X & objects of class X cannot be changed to another class \\
        X-IS-Y & objects of class X will transform to class Y \\
        X-IS-PUSH & X can be pushed \\
        X-IS-MOVE & X will autonomously move \\
        X-IS-STOP & X will prevent the player from passing through it\\
        X-IS-KILL & X will kill the player on contact\\
        X-IS-SINK & X will destroy any object on contact\\
        X-IS-[PAIR] & both rules 'X-IS-HOT' and 'X-IS-MELT' are present \\
        X,Y-IS-YOU & two distinct objects classes are controlled by the player \\
    \hline
    \end{tabular}
    \label{tab:rrp}
\end{table}

When a level is submitted to be archived, the system uses the  list of active rules at the start and the end of the level as behavior characteristic for the input level to determine its location in the map. Because there are an infinite number of rules can be created or broken in between the starting state and the win state, we decide to only focus on the rules present at the start of the level and the rules present at the end to constrain the scope of the potential levels that can be made, but also allow a freedom of solution spaces and objectives for a given starting level. There are 9 different rules checked for in each level - based on the possible rule mechanics that can be made in the Game module system. Table \ref{tab:rrp} shows the full list of possible rules. Since these rules can be active at the beginning or at the end, it makes the number of behavior characteristics equal to 18 instead of 9 which provide us with a map of $2^{18}$ cells.

The Map Module can recommend levels to start from when designing a new level. Like the Mutator Module (section~\ref{sec:mutator_module}), it also takes the Objective Module (section~\ref{sec:objective_module}) into consideration when selecting its recommendations. The Map Module can provide levels that most similarly match the objectives chosen and provide either other levels the user has previously made or high rated (and intuitively high quality) ``elite'' levels. Figure \ref{fig:select_screen} shows a screenshot of the map selection screen on the level editor page of the site where users can select from suggested levels high quality previously submitted to the site that best match a particular objective designated by the user.

In this project we are using a multi population per each cell of the Map-Elites similar to the constrained Map-Elites~\cite{khalifa2018talakat}. Levels can only be submitted to the site database and placed in the matrix if and only if they are also solvable. The quality of the level is determined by user ratings - performed by the Rating Module. By separating the levels into different cells based on the behavior characteristic of the rules present at the start and end of the level solution, we can find a large diversity of levels of varying complexity and aesthetic design. This matrix will allow a larger selection of editing and design for both the site's community and the collaborative AI system. The MAP-Elites matrix is also intended to show users what kinds of levels with particular rulespaces are missing from the database overall and guide users towards filling in these gaps while improving the quality of the levels already placed in the archive cells. This global guiding objective for the site is what makes Baba is Y'all distinct from other level editing communities - as there is a central goal for completion and improvement of the levels defined by this rulespace.

\subsection{User Profiles}

\begin{figure}[ht!]
    \centering
    \includegraphics[width=1.0\linewidth]{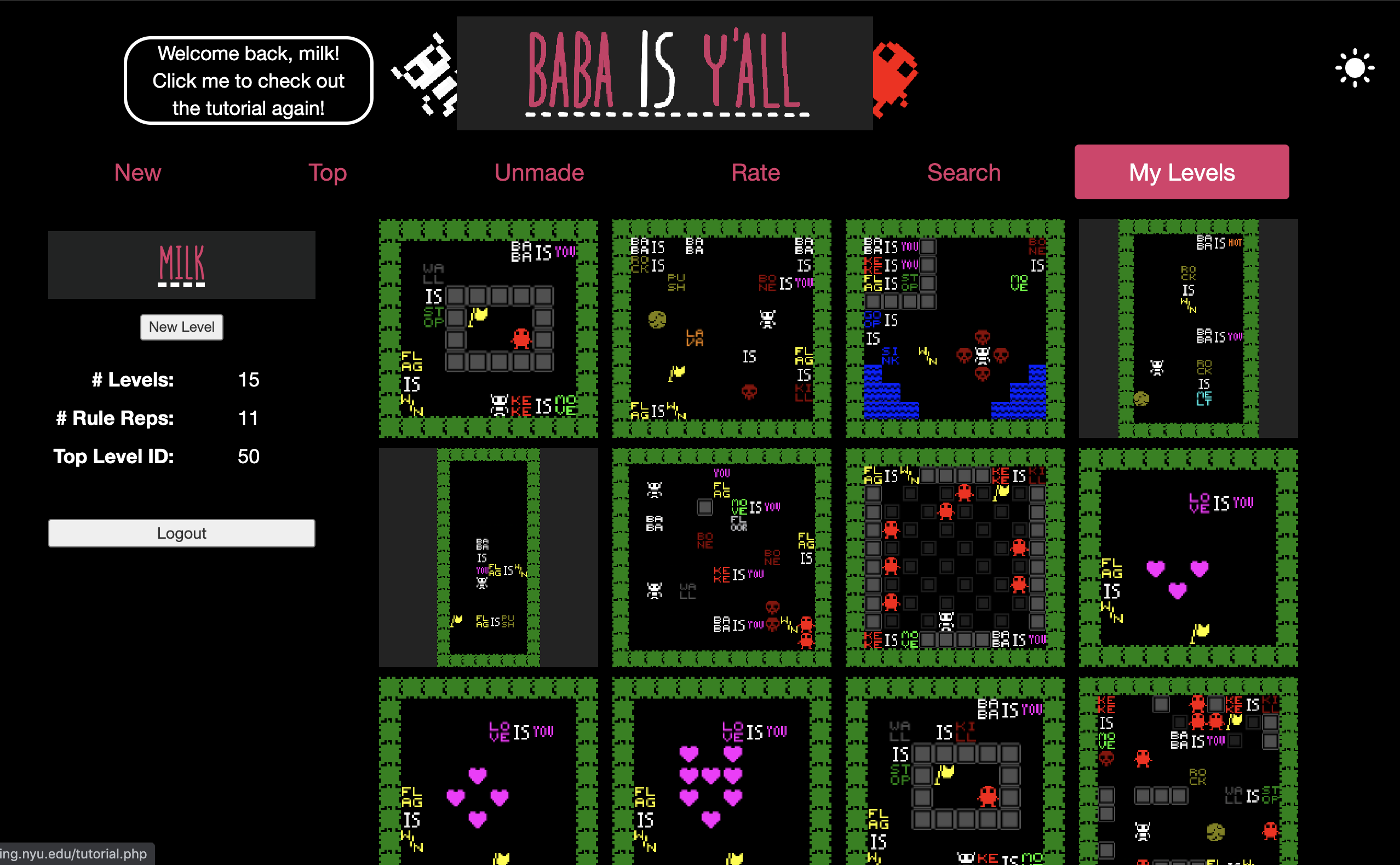}
    \caption{A screenshot of the user profile screen for the user 'Milk'} 
    \label{fig:profile_screen}
\end{figure}

The user profiles feature is the newest addition to the Baba is Y'all site. Figure \ref{fig:profile_screen} shows a screenshot of the user profile screen with the user's authored levels. Like the original system, if a user creates a profile through the site's login system and submits a level, they get authorship attributed to the submitted level. Users can also find their previously made levels on the  profile page - called ``My Levels'' - and replay them, edit them, or view the level's mechanic combination. A user's personal stats for their level submissions can also be viewed on the page including the number of levels submitted, number of rule combinations contributed, and their top rated level. This feature was implemented to provide more user agency and personalization on the site and give users better access to their own submitted levels. 

Through the search page, players can search for specific levels by username or by level name. This creates a sense of authorship over each of the levels, even if the level wasn't designed with any human input (i.e. a level with PCG.js as the author) and encourages the collaborative nature of the site between AI and human. Users may also share links to site levels via the game page.

\section{User Study Results}
To evaluate the second version of Baba is Y'all website for usability and analyze how users interact with the system as a whole, we decided to conduct a formal user study experiment. This study differs from the informal study conducted in the first version, where the results were collected exclusively from the website's level statistic data. This study gives users explicit instructions and tasks to complete and asks participants to recount a detailed report about their experience using the site, including recounting how and if they used particular features in the site, which levels they made, and their general demographic information and prior experience to using the site. This allows us to paint a better picture of what users would want from the system and how we can improve the Baba is Y'all site and design better mixed-initiative systems in the future. 

The following results were extracted from a Google Form survey given to the experiment participants. Users were instructed to play a level already made on the site, create a new level using the level editor, test it, and finally submit it to the site. They were also given the option to go through the tutorial of the site if they were unfamiliar with the `Baba is You' game or needed assistance with interacting with the level editor tool. The Google Form survey included questions asking about which features they chose to use or try to create their level (or levels) as well as some general demographic information asking about their experience with puzzle games and Sokoban-like games overall, level editors, and AI-assisted tools. This was so we could understand the general impression of what users chose to interact with and how enticing they would be to a particular designer. We also wanted to understand the familiarity and potential learning curve of a general population of people who choose to interact with level editors and AI assisting tools when designing for game levels. Those who participated in the user study were given pre-made usernames in order to verify the levels they submitted from their responses and to protect their identities. These users only had to provide an email address to register for both the site and the survey. The first 100 participants were compensated with a \$10 Amazon gift card if they accurately followed the instructions and completed the survey. Participants were invited via mailing lists and Twitter posts promoting the user study.

We received a total of $173$ responses, however, only $76$ of these responses were valid. These responses were evaluated based on cross-validation and verification between the saved level on the website and the level ID they submitted via the survey that they claimed they authored. Many of these invalid responses contained levels that either did not exist in the database or were claimed to be authored by another user already. This big difference in the numbers is due to releasing the system online with no security measures. This attracted a lot of bots that created multiple accounts so they could fill out the user survey via the link provided and therefore put random numbers as the level they submitted instead of the real levels a human user would have helped to design. The following results are taken from the self-reported subjective survey returned from those valid $76$ users.

\subsection{Demographic Data}\label{sec:demographics}





\begin{figure}[ht]
    \centering
    \includegraphics[width=1.0\linewidth]{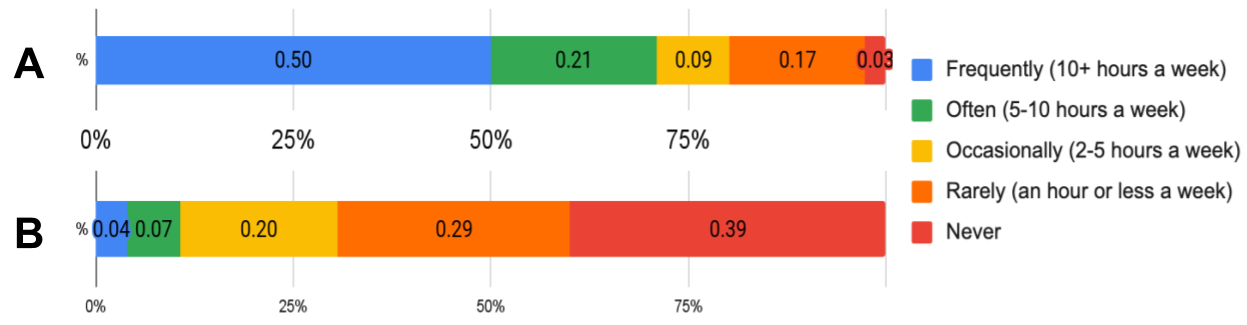}
    \caption{A. Frequency for playing games; B. Frequency for designing levels for games}
    \label{fig:freq_des_play}
\end{figure}

\begin{figure}[ht]
    \centering
    \includegraphics[width=1.0\linewidth]{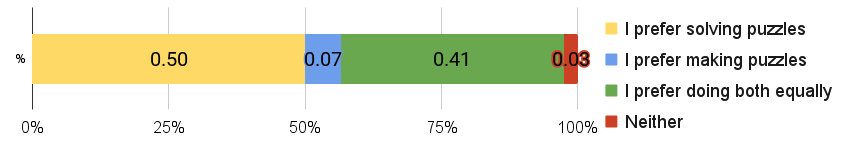}
    \caption{Preference for solving or making puzzles}
    \label{fig:design_pref}
\end{figure}

Half of the users who completed the survey answered that they frequently played video games (more than 10 hours a week) with around 80\% of the users stating they play for at least 2 hours a week (figure~\ref{fig:freq_des_play}). Conversely, only 28.9\% of users responded that they spend 2 or more hours a week designing levels for games with 40.8\% of users stating they never design levels at all (figure~\ref{fig:design_pref}). When asked if they prefer to solve or make puzzles, 50\% of participants responded that they prefer to solve puzzles, while only 6.6\% preferred the latter. 40.8\% of users were split on the preference for designing and solving puzzles. 



\begin{figure}[ht]
    \centering
    \includegraphics[width=1.0\linewidth]{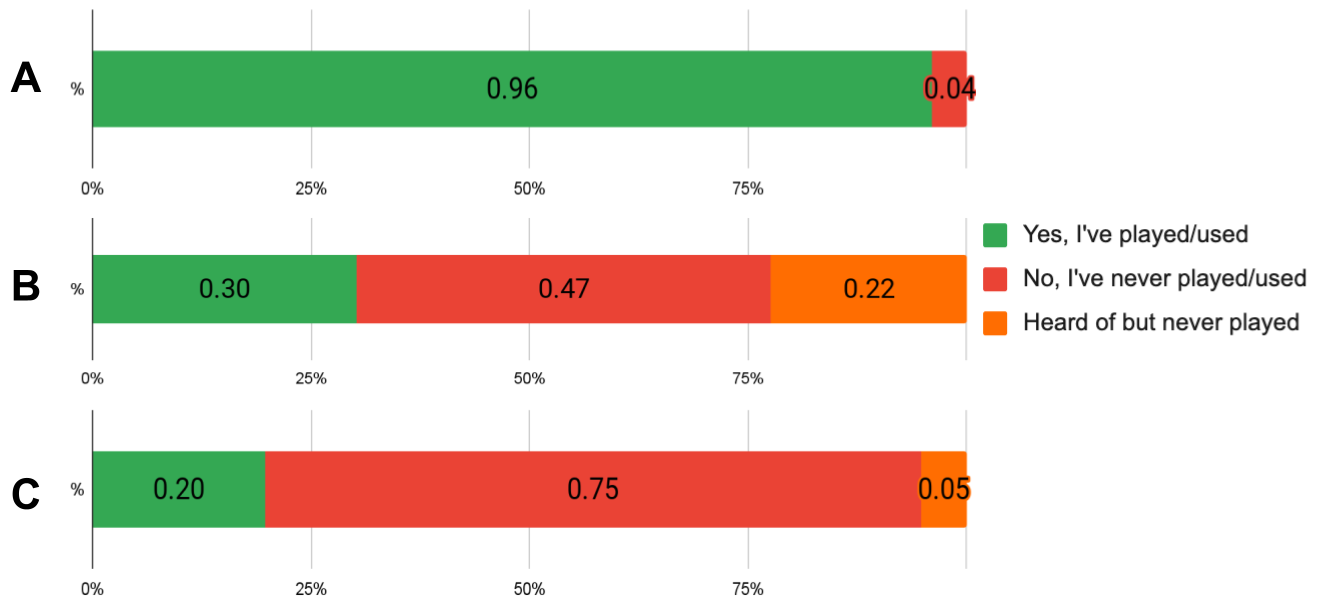}
    \caption{A. Experience playing Sokoban; B. Experience with 'Baba is You'; C. Experience with AI-assisted level editing tools}
    \label{fig:exp_graph}
\end{figure}

We asked participants if they had ever played the original game `Baba is You' by Hempuli (either the jam version or the Steam release as both contain the rules used in the Baba is Y'all site), played a Sokoban-like game (puzzle games with pushing block mechanics), and have experience with AI-assisted level editing tools. Figure~\ref{fig:exp_graph} shows the distribution of the users' answers for these questions. Only 30\% of participants had played the game before, meanwhile 22\% had heard of it but had never played it. For the rest, this study would be their first experience with the game. Interestingly enough, 96\% of the participants stated they had played a Sokoban-like game so we can infer that the learning curve would not be too harsh for the new players. Concerning AI-Assisted level editing tools, 75\% of users had never used them before, with 5.3\% stating they were unsure if they had ever used one - thus the learning curve for AI-collaboration would be much higher and new to participants.



\subsection{Self-Reported Site Interactions}

\begin{figure}[ht]
    \centering
    \includegraphics[width=1.0\linewidth]{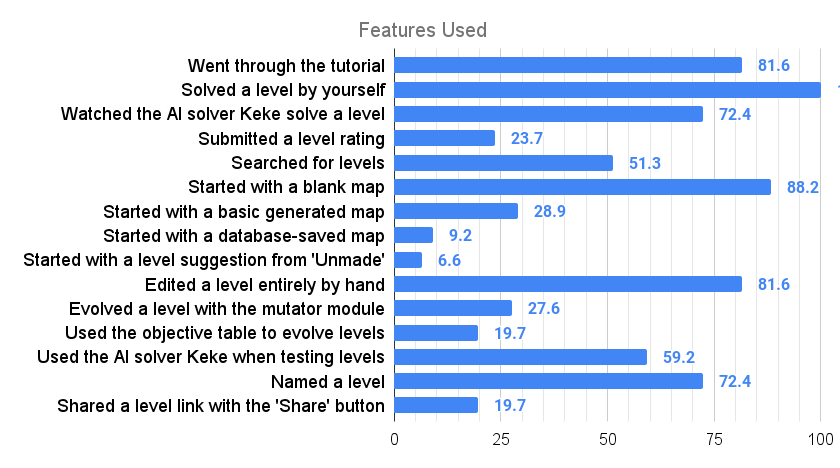}
    \caption{Survey results for users' reports on the features they used}
    \label{fig:feat_report}
\end{figure}

Figure \ref{fig:feat_report} shows the full list of features that participants interacted with on the site. Users were given the optional task to go through the tutorial section of the Baba is Y'all site to familiarize themselves with both the mechanics of the original `Baba is You' game, the AI assisted tools available to them through the level editor, and the site layout and navigation itself. 81.6\% of users went through this tutorial (whether fully or partially was not recorded.) The second task for users was to play a level that was previously submitted to the website database. 100\% of users were able to solve a level by themselves, however 72.4\% of users reported choosing to watch the Keke AI solver complete the submitted level as well. The third and final task for the participants was to submit their own `Baba is You' level using the level editor. Here, users were asked the most about their involvement with the AI system. Some users chose to create more than one level, so they may have multiple experiences and their design choices may not be mutually exclusive (i.e. using a blank level and also using an AI-suggested level.)

For the initial creation of the level, 88.2\% of users chose to start with a blank map. 9.2\% of users started with a level that had already been submitted to the level database - either a level that had been ranked as an elite level or a level created by the user themselves (in the case that they submitted more than one level during this study.) 6.6\% of users started with a level that was suggested from the 'Unmade' page - ideally with the intent to make a level with a rule combination that had not been made yet - thus expanding the MAP-Elites rule combination matrix in the database. Unfortunately, we forgot to ask users in the survey if they started with the random level option that was also provided by the AI assistance tool - so we lack data to report on this statistic. 

For editing the level, 81.6\% of users reported editing a level completely by hand without any AI assistance. 27.6\% of users edited the level with help from either the evolver algorithm or the mutator functions provided by the AI assistance back-end.  19.7\% of users reported using the objective table to aid the evolver tool in creating the level. We think this low percentage is attributed the fact that a large population of users were unfamiliar with the system or `Baba is You' game overall. This - as well as the lack of selection for level comparison from the previously submitted levels in the database - made using the evolver tool towards certain goals too steep of a task to accomplish and learn. Finally, when testing the level, 59.2\% of users reported using the Keke solver AI when testing their levels and 72.4\% of users named their levels.

While not required in the tasks given, we also asked participants about any extra site features they chose to explore. 23.7\% of users reported submitting a level rating from the 'Rate' page. 51.3\% of users reported using the 'Search' tool to search for specific levels (what their search criteria was we did not ask.) Finally, 19.7\% of users reported using the 'Share Level' to share a submitted level link with others online. 

The least used interactions - 'Started with a database-saved map in the level editor', 'Started with a level suggestion from the Unmade page', and 'Used the objectives table to evolve levels' - were also all related to the AI mixed-initiation of the system. The first could be attributed to a lack of overall levels in the database (at the start of the experiment there were only around 40 available levels) therefore leading to a lack of viable options for the user to choose from. However, the lack of usage for the other two features could be attributed to the opposite problem of having too many options to choose from - again due to lack of levels available to choose from in the database. Trying to make a level with constrained parameters may have also been too steep of a task to accomplish for someone who was totally unfamiliar with the system or even the `Baba is You' game overall. There was also no incentive for a player to create a level suggested by the system as opposed to making a level from scratch. We also didn't explicitly instruct users to make a level from the suggested set, and instead allowed them to make whatever level they wanted with the editor - whether with the prompted ruleset or from their own ideas.

\section{Website Level Statistics}




The following results reflect the statistics of the levels submitted to the Baba is Y'all website by the 76 valid users from the user study and do \textit{not} include all of the levels submitted to the site. These results also do not include the levels submitted by users outside of the study. 
The current full data breakdown of the levels submitted to the entire site can be found on the stats page of the Baba is Y'all website\footnote{url{http://equius.gil.engineering.nyu.edu/stats\_about.php}}.

\subsection{Authorship}

\begin{figure}[ht]
    \centering
    \includegraphics[width=0.95\linewidth]{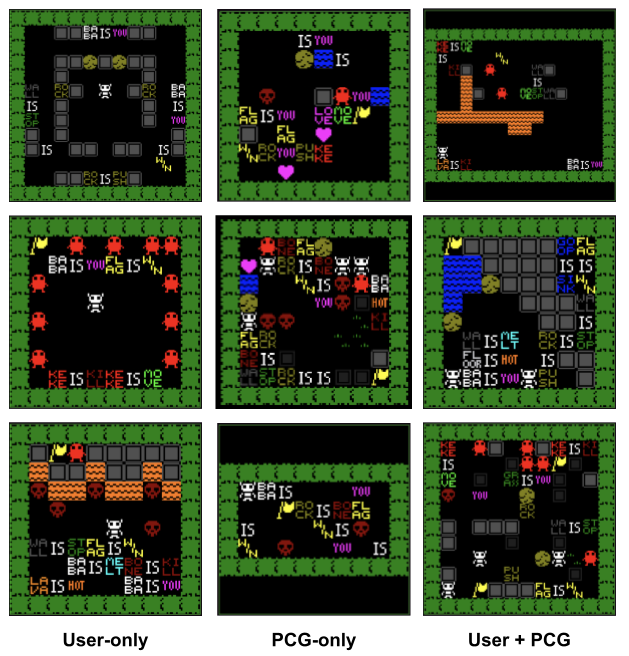}
    \caption{Sample levels generated for the system. The left column is user generated levels, the middle column is evolver module levels, and the right column is mixed-initiative user and evolver levels}
    \label{fig:level_types}
\end{figure}

We looked into all the levels created by the valid 76 users and we divided them based on how the mixed-initiative tool was used to author them. We divided them into three main categories (as shown in figure~\ref{fig:level_types}):
\begin{itemize}
    \item \textbf{User-Only levels:} were created from a blank map exclusively by the human user without any AI assistance.
    \item \textbf{PCG-only levels:} were created solely by the AI tool without any human input aside from choosing which tool to use and when.
    \item \textbf{Mixed-author levels:} involved both the human user as well as the AI tool in the creation process of the level. 
\end{itemize}


The majority of the levels submitted were user only (70.96\%), however over a quarter (26.88\%) of the levels submitted had mixed-authorship. This is a lower percentage of mixed-authorship levels than the first version of the Baba is You system - which was reported in the previous paper to have had a mixed-authorship percentage of 35/58 of the submitted levels, or 60.3\% of the levels submitted. This is a disappointing finding considering how much focus we dedicated to making a more seamless AI-human collaboration environment and trying to improve the system overall. We will have to conduct further analysis and a more in-depth study to understand why users prefer to author their levels manually using this system rather than with any assistance from the agent.

\subsection{Rule Distribution}

\begin{figure}[ht]
    \centering
    \includegraphics[width=1.0\linewidth]{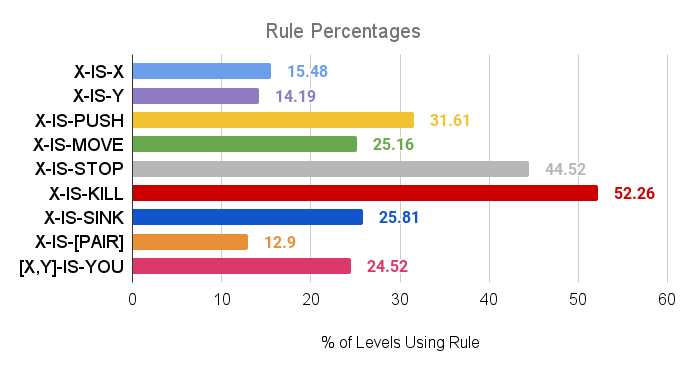}
    \caption{Site results for the rule distribution across levels submitted}
    \label{fig:level_rule_dist}
\end{figure}

From the $96$ submitted levels by the study participants, we found only $57$ different cells in the MAP-Elites matrix were covered. This is less than 1\% of the whole number of possible rule combinations ($2^{18}$ possible combinations); however it is worth noting that more than half of these submitted levels covered a distinct cell in the matrix. Figure~\ref{fig:level_rule_dist} shows the rule distributions over all of the levels submitted. The X-is-KILL rule was used the most in over half of the levels submitted and the X-is-STOP rule was used the second-most at 45.83\%. This may be because these rules create hazards for the player and add more depth to the level and solution. Meanwhile, the X-is-Y rule was used the least in only 10.50\% of the levels submitted. Its counterpart rule, X-is-X, which prevents sprites from changing form, was also used the second least amount. This rule could also have been considered the least intuitive of the rules as well - as they require an understanding of the relationship between sprite transformations.  

\begin{table}[ht]
\begin{center}
\begin{tabular}{|c c c c|} 
 \hline
 Author Type & \# Rules & Sol. Length & Map Size (\# tiles) \\
 \hline\hline
 User-only & 2.54 $\pm$ 2.24 & \textbf{28.33 $\pm$ 26.34} & 116.75 $\pm$ 45.47\\
 \hline
 PCG-only & 1.5 $\pm$ 0.5 & 32.5 $\pm$ 14.5 & 100 $\pm$ 0\\
 \hline
 Mixed-author & \textbf{2.76 $\pm$ 2.43} & 28.12 $\pm$ 22.21 & \textbf{122.52 $\pm$ 46.48}\\
 \hline
\end{tabular}
\end{center}
\caption{Averaged attributes for different types of created levels}
\label{tab:avg_author}
\end{table}

\begin{figure}[ht]
    \centering
    \includegraphics[width=1.0\linewidth]{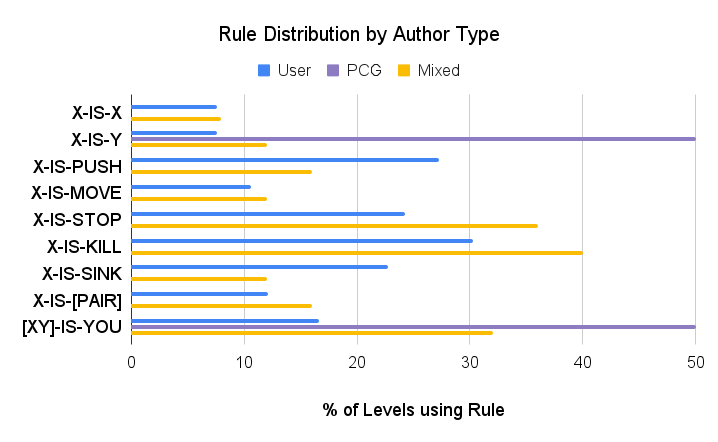}
    \caption{Rule distributions across the different authored levels}
    \label{fig:rule_dist}
\end{figure}

The relation between rules and the different type of authors can be shown in table~\ref{tab:avg_author}. Some levels may use no rules at all and only contain the required X-is-YOU and X-is-WIN rules. The mixed-author levels has the highest number of average rules per level ($2.833$), while PCG-only levels have the lowest average ($1$). The rule distributions for each author type are shown in Figure \ref{fig:rule_dist}. For 7/9 of the possible rule combinations, the Mixed-author levels had the most levels associated with for each rule - and therefore more variability overall. This could have been because of the back-end AI system trying to create more variety with rule combinations in the levels and therefore adding more level variants to the site's dataset. It is also worth noting that 24\% of the levels authored only by Users and 24\% of the levels with the Mixed-authorship had none of the 9 rules in the level. Mixed-author levels also had the highest average level size, but a close second for the solution length - just behind User-authored levels. 

\section{Questionnaire and Level Data Correlation}
\begin{table}[t!]
\resizebox{\columnwidth}{!}{%
\begin{tabular}{l|lll|lll|lll|}
\cline{2-10}
 & \multicolumn{3}{l|}{Baba is You?} & \multicolumn{3}{l|}{Puzzle Level Preference} & \multicolumn{3}{l|}{AI-Assisted Level Editors?} \\ \cline{2-10} 
 & \multicolumn{1}{l|}{Yes} & \multicolumn{1}{l|}{Maybe} & No & \multicolumn{1}{l|}{Solve} & \multicolumn{1}{l|}{Design} & Both & \multicolumn{1}{l|}{Yes} & \multicolumn{1}{l|}{Unsure} & No \\ \hline
\multicolumn{1}{|l|}{\# Mixed Levels} & \multicolumn{1}{l|}{9} & \multicolumn{1}{l|}{6} & 7 & \multicolumn{1}{l|}{12} & \multicolumn{1}{l|}{2} & 8 & \multicolumn{1}{l|}{3} & \multicolumn{1}{l|}{0} & 19 \\ \hline
\multicolumn{1}{|l|}{\# User Levels} & \multicolumn{1}{l|}{14} & \multicolumn{1}{l|}{10} & 28 & \multicolumn{1}{l|}{25} & \multicolumn{1}{l|}{5} & 22 & \multicolumn{1}{l|}{11} & \multicolumn{1}{l|}{4} & 37 \\ \hline
\multicolumn{1}{|l|}{\# PCG Levels} & \multicolumn{1}{l|}{0} & \multicolumn{1}{l|}{1} & 1 & \multicolumn{1}{l|}{1} & \multicolumn{1}{l|}{0} & 1 & \multicolumn{1}{l|}{1} & \multicolumn{1}{l|}{0} & 1 \\ \hline
\multicolumn{1}{|l|}{Avg \# Rules} & \multicolumn{1}{l|}{1.96} & \multicolumn{1}{l|}{1.53} & 1.67 & \multicolumn{1}{l|}{1.63} & \multicolumn{1}{l|}{0.57} & 2.10 & \multicolumn{1}{l|}{1.53} & \multicolumn{1}{l|}{3.75} & 1.63 \\ \hline
\multicolumn{1}{|l|}{Avg Sol. Len} & \multicolumn{1}{l|}{30.87} & \multicolumn{1}{l|}{30.65} & 28.47 & \multicolumn{1}{l|}{32.08} & \multicolumn{1}{l|}{25.14} & 27.77 & \multicolumn{1}{l|}{24.00} & \multicolumn{1}{l|}{43.50} & 30.21 \\ \hline
\end{tabular}%
}
\caption{Level statistic between selected questionnaire responses and submitted levels}
\label{tab:quest_level}
\end{table}

We merged the two datasets between the level data submitted to the website and the responses from the user study questionnaire to further analyze any correlations between answers and types of levels submitted. Table \ref{tab:quest_level} shows the raw results from 3 selected questionnaire response groups.

For the question determining the familiarity of users with the game Baba is You, of the 36 users who had never played or heard of 'Baba is You' 77\% chose to make a level that was exclusively user-authored. For the other two groups, those who had heard of the game but never played it and those who had played the game before, 62.5\% and 56\% of users respectively chose to make user-only levels. 41\% of the mixed-authorship levels submitted from the participants of the study came from the group who had played Baba is You before. From this we can infer that those who were more familiar with the game environment were more likely to use the AI assisting tool to design their levels. Users who had played the game before also had the highest average for number of rule combinations in their levels at 1.96. 

We also observed some correlations between level submissions and preference for designing or solving puzzle levels. 60\% of the participants who answered that they preferred to solve puzzle levels submitted user-only authored levels - however this group also held the majority for number of mixed-authorship levels submitted at 54.5\%. This group also had the highest solution length average of the 3 groups at an average of 32.08 steps. 28.5\% of participants who answered that they prefer to design levels submitted mixed-authorship levels to the site - the highest percentage of the 3 groups.

An interesting correlation was found from users who never had experience with AI assisted level editors before. Of the 57 users, 74\% chose to submit a user-only authored level. However, 86.3\% of the mixed-authored levels submitted from the study also came from this group. Only 20\% of users from the group who had experience with AI-assisted level editing tools chose to submit mixed-authorship levels. From this we can infer that users with little experience in AI-assisted tools are more willing to experiment with them and try to design with them.

\section{Discussion}
\subsection{Data Analysis}
It is clear from both the submitted level statistics of the site and the self-reported user survey that mixed-authorship is not the preference for users when designing levels. Many users would still prefer to have total control over their level design process from start to finish. For future work, we can look to limit user control and encourage more AI-assistance with the design process similar to the work done by Bhaumik et al.~\cite{bhaumik2021lode}. 

The limitations of the AI back-end (both the evolver and solver) may be at fault for the lack of AI interaction. The mutator and evolver system are dependent on previously submitted levels and level ratings in order to ``learn'' how to effectively evolve levels towards high quality design. As a result, the map module's ranking system is always updating its evaluation for what makes a ``good'' level based on these incoming ratings. If there is a lack of available data for the tool to learn from, the AI will be unable to create quality levels - causing the user to less likely submit mixed-initiative co-created levels, and causing a negative feedback loop. 

The fitness function defined for the evolver and mutator tool may be inadequate for level designing. It could produce a level that is deemed ``optimal'' in quality by its internal definition, but may actually be sub-par in quality for a human user. Another flaw in the AI-collaboration system, could be that the users lacked direct control on the evolver and mutator and attempting to use them in middle of creation might have been more problematic as it could destroy some of the level structures that the users were working on. Future work could remedy this problem by giving users various mutation ``options" similar to the AI selections in RLBrush \cite{delarosa2021mixed} and Pitako. \cite{machado2019pitako} Finally, the `Keke' AI solver was also lacking in performance as a few participants mentioned that the solver was unable to solve their prototype levels that they themselves could end up solving in just a couple of moves. An improved AI solver would help with the level creation efficiency.

\subsection{User Comments and Feedback}
We gave the participants opportunities to provide open feedback about their experience using the site in order to gather more subjective data about their experience as well as collect suggestions for potential new features.

Almost no users experienced any technical difficulties or bugs that prevented them from using the site. The few that did mentioned formatting issues with site caused by their browser (i.e. icons too close together, loading the helper gifs, font colors.) However, one user reported that this issue may have been because they were using the site from their phone (we unfortunately did not provide users with instructions to complete the study on a desktop or laptop.) In the future, we will be sure to exhaustively test the site on as many browsers as possible - both desktop-based and mobile - to be more accessible. 

Some users were confused by the tutorial and the amount of information it conveyed for the entire site citing it as ``intimidating'', ``overwhelming'', and ``a bit complex''. However, other users reported the lack of information saying it was ``not detailed'', or had ``sufficient information [...] but could have been delivered in a more comprehensible way.'' To make the game more accessible, we will most likely try to make the tutorial section less intimidating to new users by limiting the amount of information shown (possibly through a ``table of contents'' as suggested by one participant) while still being comprehensible enough to understand the level editor and tools. 

For feature suggestions, many users wished for larger maps and vocabulary - like those found in the Steam-release `Baba is You' game. Users also wished for a save feature that would allow them to make ``drafts'' of their level to come back later to edit. Many users also suggested a co-operative multiplayer feature for level editing and level solving - we can assume with another human and not an AI agent.

\begin{figure}[ht]
    \centering
    \includegraphics[width=1.0\linewidth]{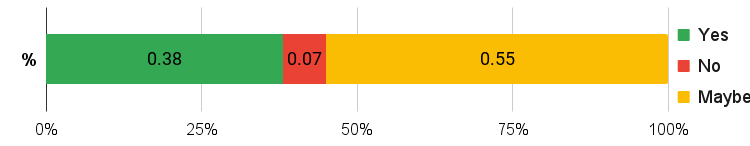}
    \caption{User feedback for likelihood to return using the site after the experiment}
    \label{fig:reuse_likelihood}
\end{figure}

While the results of the statistics on the levels submitted were disappointing for involvement of the AI assisting tool, we also asked users how likely they would continue using the site after the experiment. 38.2\% of users said they would continue to use the site, while 55.3\% said they would maybe use the site (figure~\ref{fig:reuse_likelihood}). Many users were optimistic and encouraging with the concept of incorporating AI and PCG technologies with level design - citing the project as a ``cool project'', ``a very unique experience'', a ``lovely game and experiment'', and ``very fun.'' At the time of writing, a few users did return, as their 'Keke' assigned usernames were shown as authors on the New page, long after the study was completed. Most notably, the Keke subject user Keke978 who took up the username 'Jme7' and contributed 28 more levels to the site after the study was concluded and currently holds the title for most levels submitted and most rule combinations on the site.


Many users also provided us with constructive feedback for feature implementation, site usability, and suggestions for improvement with how to further incorporate the AI back-end interactivity. As shown in figure~\ref{fig:exp_graph}, 70\% of users who played with the system had never played the game `Baba is You' and 75\% of people had never used an AI-assisted level editor tool before this experiment. Based on this information and retainability of users to complete the survey and provide the constructive feedback, we can extrapolate 2 conclusions: 1. the game stands alone, independent of `Baba is You', as an entertainment system; and 2. for people with even limited AI-gaming experience, as long as they are not completely foreign to gaming, this project has the ability to grasp their attention long enough to understand it, tinker around, and then give constructive feedback.
 

\section{Conclusion and Future Work}
The results from the user study have demonstrated both the benefits and limitations of a crowd-sourced mixed-initiative collaborative AI system. Currently, users still prefer to edit most of the content themselves, with minimal AI input - due to the lack of submitted content and ratings for the AI to learn from. Pretraining the AI system before incorporating it into the full system would be recommended to create more intelligent systems that can effectively collaborate with their human partners for designing and editing content. This would lead to more helpful suggestions on the evolver's end as well as better designed levels overall. This project is the start of a much longer and bigger investigation into the concept of crowd-sourced mixed initiative systems that can use quality diversity methods to produce content and we have many more ideas to improve upon the Baba is Y’all system. 

As suggested by many participants in the user study, we would like to incorporate level design collaborations between multiple users and multiple types of evolutionary algorithms all at once to create levels. Our system would take inspiration from collaboration tools such as LodeEncoder \cite{bhaumik2021lode}, RLBrush \cite{delarosa2021mixed}, and Roblox (Roblox Corporation, 2006). This would broaden the scope and possibilities of level design and development even further to allow more creativity and evolutionary progress within the system. This collaboration setting will open multitude of interesting problems to investigate such as authorship. 

Outside of the `Baba is You' game, we would like to propose the development of an open-source framework to allow mixed-initiative crowd-sourcing level design for any game or game clone. Such games could include Zelda, Pacman, Final Fantasy, Kirby, or any other game as long as we have a way to differentiate between levels mechanically and we can measure minimum viable quality of levels. Adding more games to the mixed-initiative framework would allow an easier barrier of entry to players who may have been unfamiliar with the independent game `Baba is You' but is very familiar with triple-A games produced by companies such as Nintendo.

We would like to also propose a competition for the online `Keke' solver algorithm for the challenging levels. In this competition, users would submit their own agent that can solve the user-made and artificially created `Baba is You' levels. Ideally, this improve the solver of the `Baba is Y'all' system but also introduce a novel agent capable of solving levels with dynamically changing content and rules - an area that has not been previously explored in the field. Development for this framework for this competition has already begun at the time of writing this paper.

Finally, we would like to propose the creation of a fully autonomous level generator and solver that can act as a user to our system. 
This generator-solver pair would work parallel to the current system's mixed-initiative approach, but with a focus on coverage to exhaustively find and create levels for every combination of mechanics. With a redefined fitness function and updated solver (possibly from the Keke Solver Competition,) this could be more efficient than having users manually submit the levels, while still using content created by human users to maintain the mixed-initiative approach. 

The second version of the Baba is Y'all system paired with the user study we conducted gave us many insights into designing a mixed-initiative creative collaboration system. First, we believe that the system is not quite at the point where an AI could match a user in terms of creative design to the point where a user would see the system as its creative equal and be willing to share control - as noted by the majority of user-exclusively authored levels submitted to the database. Second, we would like to find more ways to encourage users to want to work with the AI tool, so that the system can learn and improve to reach the point of creative equality. Thirdly, we would want to design a system that is both more transparent about what processes and decisions the AI makes when helping to design content while preventing a user's creative vision from being lost to the AI's edits and design choices. Finally, while the user study was helpful in observing what kinds of interactions a user might partake when designing within the space of this system, a more extensive study could be done to discover where the limitations and failings of the mixed-initiative collaborative system exist and how to improve on them; potentially by conducting a study over a longer period of time. There are many new directions we can take the Baba is Y'all system and the concept of crowd-sourced collaborative mixed-initiative level design as a whole and this project will hopefully serve as a stepping stone into the area and provide insight on how AI and users can work together in a crowd-sourced website to generate new and creative content.

\section*{Acknowledgments}
The authors would like to thank the Game Innovation Lab, Rodrigo Canaan, Mike Cook, and Jack Buckley for their feedback on the site in its beta version as well as the numerous users who participated in the study and left feedback.

\ifCLASSOPTIONcaptionsoff
  \newpage
\fi

\bibliographystyle{IEEEtran}
\bibliography{references}

\end{document}